\documentclass[twocolumn,showpacs,floats,amsmath,amssymb,nofootinbib]{revtex4}
\usepackage{graphicx}
\usepackage{dcolumn}
\usepackage{bm}
\usepackage{amsmath}
\usepackage{float}
\usepackage{color}
\usepackage{amsfonts}
\usepackage{amssymb}
\usepackage[all]{xy}
\begin{document}
\title{A holographic cut-off inspired in the apparent horizon}

\author{Miguel Cruz}
\altaffiliation{miguelcruz02@uv.mx}
\affiliation{Facultad de F\'\i sica, Universidad Veracruzana 91000, Xalapa, Veracruz, M\'exico}

\author{Samuel Lepe}
\altaffiliation{samuel.lepe@pucv.cl}
\affiliation{Instituto de F\'{\i}sica, Facultad de Ciencias, Pontificia Universidad Cat\'olica de Valpara\'\i so, Avenida Brasil 4950, Valpara\'\i so, Chile\\}

\date{\today}

\begin{abstract}
In this work we consider a general function for the $c^{2}$ term that appears in the conventional expression for the holographic dark energy in a FLRW curved spacetime. The cut-off prospective is inspired in the apparent horizon length. By exploring the slowly varying condition for the $c^{2}$ term, we obtain a range of validity for this holographic proposal. Under these considerations can be found that this holographic cut-off is adequate to describe the late cosmic evolution. Additionally, by considering some values constrained with the use of observational data for some cosmological parameters in the context of dark energy models, such as the deceleration parameter or the parameter state, can be shown that this holographic model remains {\it close} to the $\Lambda$CDM model. 
\end{abstract}


\maketitle

\section{Introduction}\label{intro}

Generally the followed path to give a description for the late cosmic evolution involves going beyond the theory of general relativity, several of the proposed models can describe an accelerated cosmic phase, however the results are not conclusive, a difficult test to pass for many of these cosmological models is to keep a small growth of the matter-density perturbations during the expansion. For an interesting review on the topic of modified theories of gravity, see for instance the Ref. \cite{clifton}. \\

In the standard framework, the current cosmic acceleration is due to an exotic component of the universe known as dark energy, this idea is not ruled out by cosmic data since a certain tendency can be observed in the value obtained for the parameter state that describes the dark energy, this value oscillates around -1 \cite{data, des, riess, perlmutter}, it is even possible to see that some results place this value below -1. The nature of the dark energy has been discussed in the literature as quintessence matter or phantom one \cite{general, libro}, but some of its fundamental properties are still an open subject.\\
     
In this paper we will approach the problem of dark energy from a holographic point of view. The concept of holography at cosmological level began to acquire interest since it could be determined that it alleviated the coincidence problem even for non-spatially flat cosmologies \cite{pavon1}, under this description the physical quantities inside the universe can be described by quantities defined on the boundary of the universe, specifically, the density associated to the dark energy must be proportional to an infrared cut-off with a particular interest in the cosmological context. See for instance the Refs. \cite{od1, od2}, where was shown that under a specific election for the horizon length of the holographic dark energy, some important consequences can be obtained at cosmological level, such as crossing of the phantom divide, unification of early time inflationary epoch and late time accelerating universe, among other. Within the holographic description is common to find interacting schemes for the dark energy with other components of the universe \cite{int}, providing a natural framework for thermodynamics description, however, some issues at thermodynamics level for an expanding universe deserve a deeper investigation \cite{chemical}.\\  

In this work we will explore the effects of the curvature parameter on the resulting cosmological quantities for a specific holographic proposal, which is inspired in the length of the apparent horizon, as we will see later, the $c^{2}$ term appearing in the holographic dark energy is a general function of the redshift and it is written in terms of the curvature parameter. For a flat universe case the $c^{2}$ term becomes a constant, this resembles the Li holographic model \cite{li}.\\
Besides, by evaluating the $c^{2}$ term at present time it is possible to establish a range of validity for an arbitrary constant that appears in the holographic cut-off for the dark energy density, as expected, the value of this constant depends on the value of the curvature parameter and an important consequence is that such range of values, is appropriate to describe the late time cosmic evolution in this model, i.e., an expanding universe.\\
 
The curvature parameter seems to have a relevant role in the holographic scheme, in Ref. \cite{joel} can be found that for a closed universe the interacting term for the dark cosmological sector depends on this value. On the other hand, in Ref. \cite{miguel} was found that only for a non-flat universe, the interacting description for the dark energy - dark matter components admits a future singularity. As far as we know, a non-flat universe is not ruled out by the observational data \cite{data} and some recent results coming from the study of the Hubble parameter for several dark energy models with spatial curvature contribution showed that the $\Lambda$CDM model does not discard this kind of models \cite{curv1, curv2}.\\ 

Finally, for this holographic model it can be established that it is appropriate to describe a late cosmic evolution and according to observations, the coincidence parameter is of order 1 at present time, later, as the universe evolves to the far future, the coincidence parameter decreases.\\
   
This work is organized as follows: In Section \ref{sec:dynamics} we describe the dynamics of the model in a FLRW curved spacetime and we provide some generalities of the apparent horizon inspired holographic cut-off used in this approach. In Section \ref{sec:slow} we study the slow varying condition for the $c^{2}$ term in this holographic proposal, we establish the stage of the cosmic evolution that this holographic model can describe. In Section \ref{sec:omega} we study some consequences on the parameter state of the model, we determine some values of the cosmological parameters of the model at present time. In Section \ref{sec:final} we write the final remarks of our work.

\section{Dynamics of the model}
\label{sec:dynamics}

For two non-interacting fluids characterized by densities $\rho_{1}$ and $\rho_{2}$, the Friedmann constraint in a curved FLRW spacetime can be written as follows
\begin{equation}
3H^{2}\left(1-\Omega_{k}\right) = \rho_{1}\left(1+r\right),
\label{eq:Fried}
\end{equation}
where $r$ is the coincidence parameter defined as $r := \rho_{2}/\rho_{1}$ and $\Omega_{k}$ is the curvature parameter given by $\Omega_{k}(z) = - k (1/a_{0}^{2}H_{0}^{2})(1+z)^{2} = \Omega_{k}(0)(1+z)^{2}$, being $a_{0}$ the value of the scale factor at present time ($z=0$), $H_{0}$ the Hubble constant (value of the Hubble parameter for $z=0$), $k = \pm 1 , 0$, represents a closed, open and flat universe, respectively and $z$ is the redshift. In this case $\rho_{1}$ is the density for the dark energy component and as we will see in the next section we will consider a holographic cut-off inspired in the apparent horizon length to describe it and $\rho_{2}$ the corresponding one for the dark matter component.\\ 

The continuity equations for the energy densities have the following expressions
\begin{align}
& \rho'_{1} - 3\frac{(1+\omega_{1})}{(1+z)}\rho_{1} = 0, \label{eq:cont1}\\
& (r\rho_{1})' - 3\frac{(1+\omega_{2})}{(1+z)}r\rho_{1} = 0,
\label{eq:cont2} 
\end{align}   
where the prime stands for redshift derivative and $\omega_{1,2}$ are the parameters state which relate the energy density of the fluid with its pressure. %By considering the Eqs. (\ref{eq:Fried}), (\ref{eq:cont1}) and (\ref{eq:cont2}), together with the definition of the curvature parameter, $\Omega_{k}$, the deceleration parameter takes the form
%\begin{eqnarray}
%1+q(z) &=& (1+z)\frac{d \ln H(z)}{dz}, \nonumber \\
%&=& \frac{1}{2}\left\lbrace \frac{3(1+\omega_{1})+3r(1+\omega_{2})}{1+r}+\frac{2\Omega_{k}}{1-\Omega_{k}}\right\rbrace.
%\end{eqnarray}
%It is important to note that the curvature parameter has a direct incidence on the deceleration parameter %for this model.

\subsection{Inspired apparent horizon cut-off}
Based on the holographic principle we deal with $\rho \sim L^{-2}$, where $L$ is the size of the current universe. From now on, we will consider the energy density for the dark energy as follows
\begin{equation}
\rho_{1} = 3\left(\beta_{1}-\beta_{2}\Omega_{k}\right)H^{2},
\label{eq:density}
\end{equation}
where $\beta_{1}$ and $\beta_{2}$ are constant parameters and the factor 3 is introduced for convenience. The above expression has a similitude  with apparent horizon radius defined as $r_{ah} = \left[(1-\Omega_{k})H^{2}\right]^{-1/2}$. It is important to point out that the holography based on the use of the apparent horizon obeys the first law of thermodynamics and naturally a gravitational entropy can be associated with the apparent horizon \cite{bak}. In terms of the redshift the energy density (\ref{eq:density}) is simply
\begin{equation}
\rho_{1}(z) = 3\left[\beta_{1}- \beta_{2}\Omega_{k}(0)(1+z)^{2}\right]H^{2}(z),
\label{eq:density1}
\end{equation}
note that $\rho_{1}(z\rightarrow -1) \rightarrow 3\beta_{1}H^{2}$. As can be seen from the expression (\ref{eq:density1}), the factor of $H^{2}$ is a function of the redshift, this differs from the proposal made by Li in Ref. \cite{li}, where this factor is only a constant given by $3c^{2}$. However, in Ref. \cite{pavon} can be found that generally the $c^{2}$ term is considered as a constant in the interval $0 < c^{2} < 1$. According to its value, this constant has an important role since can provide a cosmological constant cosmic expansion or the corresponding one to an eternal expansion. For a more general description the $c^{2}$ term must be assumed as a slowly varying function of time, in this sense from our previous equations we get
\begin{equation}
\rho_{1}(z) = 3c^{2}(z)H^{2}(z), 
\label{eq:rhored}
\end{equation}
where the following equivalence can be established
\begin{equation}
c^{2}(z) := \beta_{1}-\beta_{2}\Omega_{k}(0)(1+z)^{2}.
\label{eq:ccuad}
\end{equation}    
From above expression we can see that at present time ($z=0$) the function $c^{2}(0) = \beta_{1}$ only for a flat universe ($\Omega_{k}(0) = 0$). If we replace the redefinition of $\rho_{1}(z)$ expressed in (\ref{eq:rhored}) into the Eq. (\ref{eq:Fried}) we have
\begin{equation}
c^{2}(z) = \frac{1-\Omega_{k}(z)}{1+r(z)}.
\label{eq:ccurvature}
\end{equation}
Note that the value of $c^{2}(z)$ at present time can be determined by two parameters known from observational data, which are the curvature and coincidence parameters \cite{data}, we obtain $c^{2}(0)$ in the interval $(0.72, 0.76)$.\\

Additionally, notice that Eq. (\ref{eq:ccurvature}) is in agreement with the interval $0 < c^{2} < 1$. It is worthy to mention that the Eq. (\ref{eq:ccuad}) can be written in terms of its value at present time as follows
\begin{equation}
c^{2}(z) = \beta_{1}+\left[c^{2}(0)-\beta_{1} \right](1+z)^{2},
\label{eq:function}
\end{equation}
from last expression we can see that the function $c^{2}(z)$ has only one free parameter, i.e., $\beta_{1}$, therefore
\begin{equation}
\rho_{1}(z) = 3\left\lbrace \beta_{1}+\left[c^{2}(0)-\beta_{1} \right](1+z)^{2}\right\rbrace H^{2}(z).
\end{equation}

\section{The $c^{2}(z)$ function}
\label{sec:slow}

In this section we will establish a relation between the constant $\beta_{1}$ coming from the holographic cut-off and the present time value of the $c^{2}$ term by means of the slow variation condition. As commented previously, the function $c^{2}$ must be a slowly varying function, this means that is bounded by the Hubble expansion rate, $H$, i.e., \cite{pavon}
\begin{equation}
\frac{\left[c^{2}(t)\right]^{\dot{}}}{c^{2}(t)} \lesssim H,
\end{equation}
where the dot represents a cosmic time derivative, note that this condition must be guaranteed at all times. In our case the previous equation is simply
\begin{equation}
(1+z)\frac{1}{c^{2}(z)}\frac{d c^{2}(z)}{dz} \gtrsim -1,
\label{eq:condition}
\end{equation}
where the redshift and scale factor are related through the expression $1+z = a_{0}/a$. By considering the expression for $c^{2}(z)$ given in Eq. (\ref{eq:function}) in the above equation, we obtain the following condition
\begin{equation}
-3\left[c^{2}(0)-\beta_{1} \right]\left(1+z\right)^{2} \lesssim \beta_{1},
\label{eq:betacond}
\end{equation}
which at present time results
\begin{equation}
\beta_{1} \lesssim \frac{3}{2}c^{2}(0).
\label{eq:slow}
\end{equation} 
Note that for all $z$ we have the condition $c^{2}(z) \gtrsim (2\beta_{1})/3$, therefore $(2\beta_{1})/3 \lesssim c^{2}(z) < 1$. However, since the condition (\ref{eq:condition}) must be preserved along the cosmic evolution, we can write
\begin{equation}
1+z \lesssim \sqrt{\frac{\beta_{1}}{3\left[\beta_{1}-c^{2}(0)\right]}}.
\label{eq:redsim}
\end{equation}
From the previous Eq. we can infer that $\beta_{1} > c^{2}(0)$ in order to have a real valued expression. Therefore, from this last condition and the Eq. (\ref{eq:slow}) we obtain
\begin{equation}
c^{2} < \beta_{1} \lesssim \frac{3}{2}c^{2}(0).
\end{equation}
As can be seen from the Eq. (\ref{eq:redsim}), as we approach the far future $(z=-1)$, the condition of slow variation imposes the following result, $\beta_{1} \gtrsim 0$, therefore $0 \lesssim \beta_{1} \lesssim (3c^{2}(0))/2$.\\

On the other hand, by considering the Eq. (\ref{eq:slow}) in the Eq. (\ref{eq:redsim}), one gets that $z \lesssim 0$, this is a clear indication that the suggested holographic cut-off for the dark energy density is appropriate to describe the late cosmic evolution.\\

\begin{figure}[htbp!]
\centering
\includegraphics[scale=0.65]{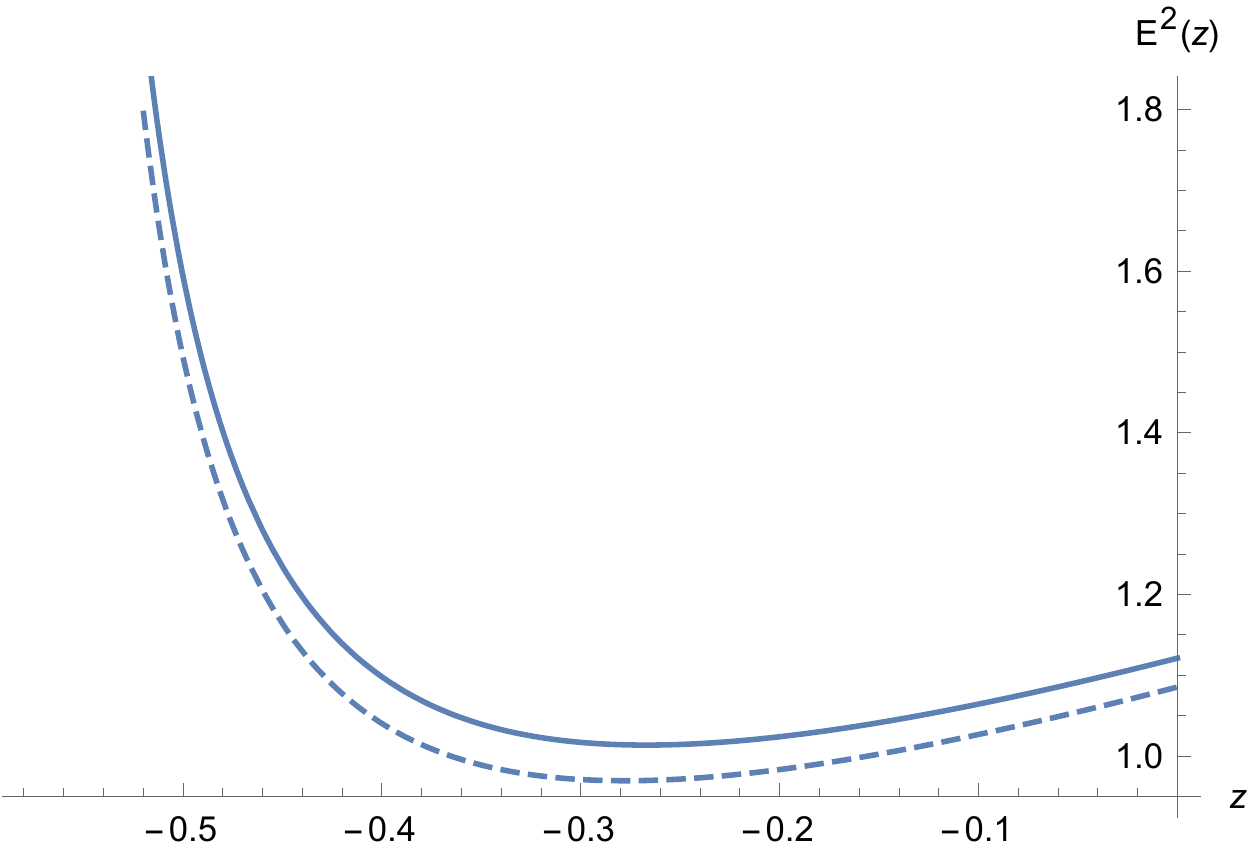}
\caption{Normalized Hubble parameter as a function of the redshift. The solid line is for $\Omega_{k}(0)= 0.005 (k=-1)$ and the dashed line represents the case $\Omega_{k}(0)= -0.005 (k=1)$.} 
\label{fig:normalized}
\end{figure}

From the Friedmann constraint given in Eq. (\ref{eq:Fried}), we can compute the normalized Hubble parameter, yielding
\begin{eqnarray}
E^{2}(z) &=& \frac{\Omega_{2}(0)(1+z)^{3}+\Omega_{k}(0)(1+z)^{2}}{1-c^{2}(z)}, \nonumber \\
&=& \frac{\Omega_{2}(0)(1+z)^{3}+\Omega_{k}(0)(1+z)^{2}}{1-\left[\beta_{1}-\left(\beta_{1}-c^{2}(0)\right)(1+z)^{2}\right]},
\label{eq:normal}
\end{eqnarray} 
where $E(z) := H(z)/H(0)$, note the dependence on the curvature parameter of the previous equation, also through the value of $c^{2}(0)$. In FIG. \ref{fig:normalized} we can observe the behavior of Eq. (\ref{eq:normal}) for a late time evolution. For this plot we have considered the interval $(0.72, 0.76)$ for $c^{2}(0)$ and the condition given in (\ref{eq:slow}) for $\beta_{1}$ and the values $\Omega_{2}(0) = 0.3089 \pm 0.0062$, $\Omega_{k}(0) = 0.000 \pm 0.005$ \cite{data}. As observed, as the model approaches the far future, the normalized Hubble parameter increases, this differs from the behavior observed in the $\Lambda$CDM model, where the Hubble parameter tends to a bounded value.
%%%%%%%%%%%%%%%%%%%%%%%%%%%%%%%%%%%%%%%%%%%%%%%%%%%%%
%%%%%%%%%%%%%%%%%%%%%%%%%%%%%%%%%%%%%%%%%%%%%%%%%%%%
\\

On the other hand, by considering the ratio between the densities $\rho_{2}$ and $\rho_{1}$ we can obtain an specific expression for the coincidence parameter
\begin{equation}
r(z) = \frac{\rho_{2}(z)}{\rho_{1}(z)} = \Omega_{2}(0)\frac{(1+z)^{3}}{c^{2}(z)E^{2}(z)},
\label{eq:coinc}
\end{equation}
where the functions $c^{2}(z)$ and $E^{2}(z)$ are known. By substituting $E^{2}(z)$ in the previous expression one gets
\begin{equation}
r(z) = \left(\frac{1-c^{2}(z)}{c^{2}(z)} \right)\left[1+\frac{\Omega_{k}(0)}{\Omega_{2}(0)(1+z)} \right]^{-1},
\label{eq:coincidence}
\end{equation}
from the previous equation we can see that $r(z) < 1$ and $r(z\rightarrow -1) \rightarrow 0$. At present time we can obtain the following constraint from the condition $r(0) < 1$
\begin{equation}
\frac{1}{2} < c^{2}(0) < 1,
\end{equation}
given that $1+\Omega_{k}(0)/\Omega_{2}(0) > 0$. Note that this is consistent with the interval $(0.72, 0.76)$ given before for $c^{2}(0)$.\\ 

\begin{figure}[htbp!]
\centering
\includegraphics[scale=0.65]{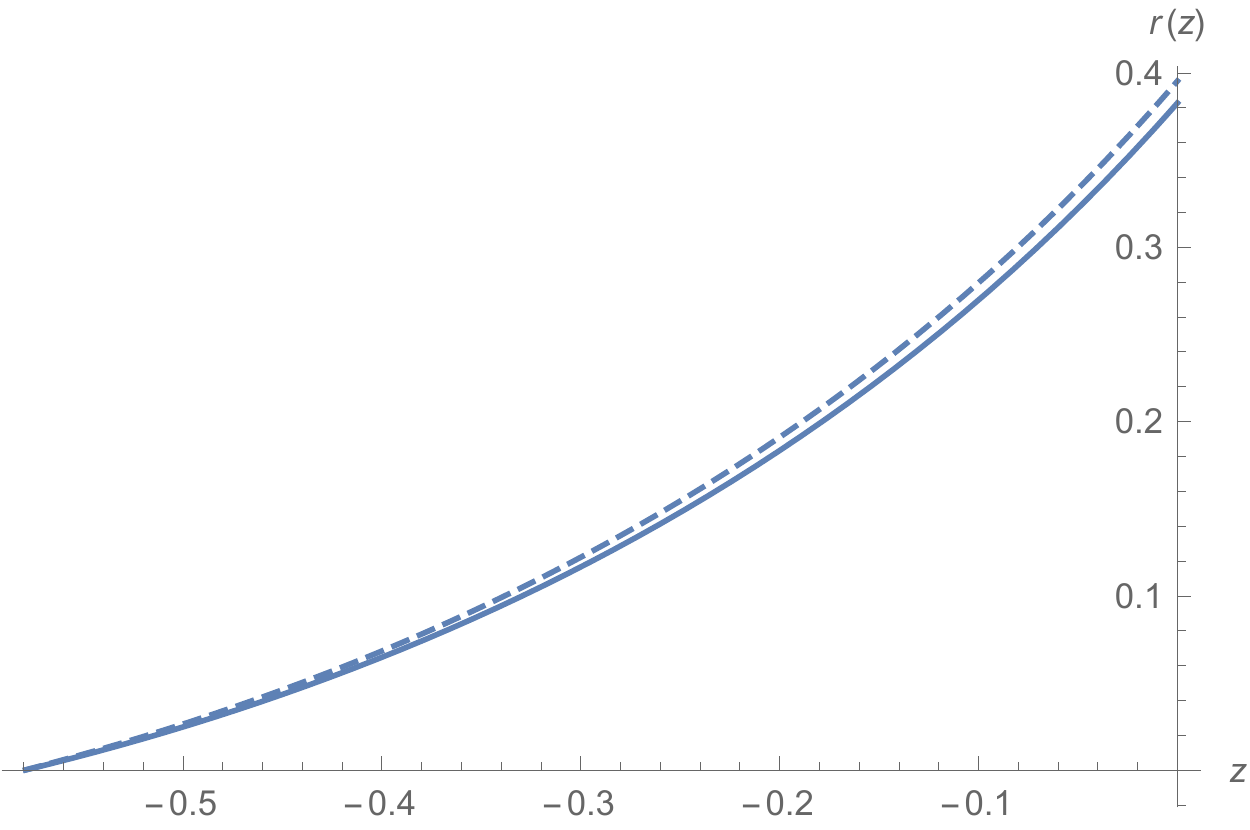}
\caption{Behavior of the coincidence parameter. The solid line is for $\Omega_{k}(0)= 0.005 (k=-1)$ and the dashed line represents the case $\Omega_{k}(0)= -0.005 (k=1)$.} 
\label{fig:coincidence}
\end{figure}

In FIG. \ref{fig:coincidence} we show the behavior of the coincidence parameter given in Eq. (\ref{eq:coinc}) as a function of the redshift, for this plot we used the values given previously for $c^{2}(0)$, $\Omega_{2}(0)$ and $\Omega_{k}(0)$. 
%%%%%%%%%%%%%%%%%%%%%%%%%%%%%%%%%%%%%%%%%%%%%%%%%%%%%
%%%%%%%%%%%%%%%%%%%%%%%%%%%%%%%%%%%%%%%%%%%%%%%%%%%%
\\

As can be observed in FIG. \ref{fig:coincidence}, the density ratio is of order one at present time, which means that both densities are of the same order of magnitude, this is known as the cosmological coincidence problem and it is an indication that we are in a special period of cosmic evolution, this fact is corroborated by observational data. However, as the universe evolves the coincidence parameter decreases, as can be seen from Eq. (\ref{eq:coincidence}), this behavior is independent of the value of the curvature and $\Omega_{2}(0)$ parameters, this is a clear indication that the dark energy dominance starts {\it now} and will keep that way, this is also a known feature of the cosmological coincidence problem \cite{velten}. 

\section{The parameter state $\omega_{1}(z)$}
\label{sec:omega}

In this section we will explore some consequences on the parameter state for the dark energy component, $\omega_{1}$. From the continuity equation given in the expression (\ref{eq:cont1}), we can write
\begin{equation}
1+\omega_{1}(z) = \frac{1}{3}(1+z)\frac{\rho'_{1}}{\rho_{1}},
\end{equation}
taking into account the Eqs. (\ref{eq:rhored}) and (\ref{eq:function}) in the previous expression, one gets
\begin{equation}
\omega_{1}(z)-\frac{1}{3} = \frac{2}{3}\left\lbrace q(z)-\frac{\beta_{1}}{\beta_{1}+\left[c^{2}(0)-\beta_{1}\right](1+z)^{2}}\right\rbrace,
\label{eq:untercio}
\end{equation}
being $q(z)$ the deceleration parameter. If we evaluate at present time the constant $\beta_{1}$ takes the form
\begin{equation}
\beta_{1} = \frac{c^{2}(0)}{2}\left\lbrace 1-3\omega_{1}(0)+2q(0)\right\rbrace,
\label{eq:beta}
\end{equation}
by considering $q(0) \sim -0.5$ and $\beta_{1} \approx 3c^{2}(0)/2$, which is in agreement with the condition given in the Eq. (\ref{eq:slow}), we obtain $\omega_{1}(0) \sim -1$, which resembles a cosmological constant evolution.\\ 

From the expression (\ref{eq:beta}) and the slow-variation result for $\beta_{1}$ given in (\ref{eq:slow}), we can provide for the deceleration parameter
\begin{equation}
q(0) \lesssim 1+\frac{3}{2}\omega_{1}(0).
\label{eq:decel}
\end{equation}
Additionally, we can write for the energy density of the dark energy the following expression
\begin{equation}
\rho_{1}(z) = \frac{3c^{2}(0)}{2}\left[3-(1+z)^{2}\right]H^{2}(z),
\end{equation}
where the slow variation condition (\ref{eq:slow}) is considered. It is worthy to mention that energy density expressed before depends on the value of the curvature parameter at present time since the constant $c^{2}(0)$ comes from Eq. (\ref{eq:ccurvature}).\\ 

In Ref. \cite{des} a dark energy equation of state was found in the interval $\omega\left( 0\right) = -0.95_{-0.39}^{+0.33}$, from the dark energy survey. If we take into account the Eq. (\ref{eq:decel}) and the aforementioned values for the parameter state, we can find that the deceleration parameter must obey the following condition $-1.01 \leqslant q(0) \leqslant 0.07$. It is important to point out that the interval obtained for the deceleration parameter includes the $\Lambda$CDM model.\\ 

In Ref. \cite{datadecel} can be found the interval $- 0.8 \leqslant q(0) \leqslant -0.2$ for the deceleration parameter when a generalized parametrization of this cosmological quantity is considered together with an updated compilation of the Hubble measurements obtained from the cosmic chronometer, with this interval for $q(0)$ and the Eq. (\ref{eq:decel}) one gets $-1.2 \leqslant \omega_{1}(0) \leqslant -0.8$, then this holographic proposal for the dark energy can behave as phantom or quintessence.

\section{Final remarks}
\label{sec:final}

By considering the apparent horizon length for the holographic dark energy, we explore the $c^{2}$ term given by a general function of the redshift. Since we are taking into account the spatial curvature of the spacetime the resulting $c^{2}$ term depends on the present time values of the curvature and coincidence parameters. For a flat spacetime at present time the $c^{2}$ function becomes a constant.\\

The form obtained for the function $c^{2}(z)$ in this holographic cut-off seems to be reasonable, by considering a slow time variation for it, or in other words, that its variation is not much faster than the scale factor, we can obtain a range of validity for this holographic proposal. At first glance the slow variation can be guaranteed in the far future and is also fulfilled in the past, therefore this holographic proposal for dark energy could be adequate to describe the cosmic evolution. The behavior of the normalized Hubble parameter in this model differs from the one obtained in the $\Lambda$CDM as the universe approaches the far future, in this model its value increases. Additionally, when the coincidence parameter is computed, at present time we can observe that is around one, which is in agreement with observational data.\\

Once that the slowly variation is considered, the parameter state for the dark energy can be obtained from the dynamics of the model. This parameter state results to be a varying function of the redshift and depends explicitly on the deceleration parameter, which is also a function of the redshift. Given that the parameter state can vary, this holographic model could mimic the $\Lambda$CDM model.\\

As a simple verification we can observe that by considering the constrained value of $\omega$ for some dark energy models at present time, we obtain that the deceleration parameter can take values below $-1$, i.e., beyond the cosmological constant cosmic evolution. As the value of the parameter state reaches the phantom zone, the deceleration parameter becomes more negative. This appears to be consistent. A cosmological constant evolution can be emulated in this model. Finally, if we consider the value constrained for the deceleration parameter with the use of recent data, we obtain that the parameter state of this holographic proposal can take values in the phantom or quintessence zone. In this work we are not considering gravitational interaction in the dark sector, we will discuss this elsewhere.

\section*{Acknowledgments}
M. C. acknowledges the hospitality of the Instituto de F\'\i sica of Pontificia Universidad Cat\'olica de Valpara\'\i so, Chile, where part of this work was done. M.C. work is supported by S.N.I. (CONACyT-M\'exico) and PRODEP (UV-PTC-851).


\begin{thebibliography}{99}
\bibitem{clifton}
T.~Clifton, P.~G.~Ferreira, A.~Padilla and C.~Skordis, Phys.\ Rept. {\bf 513}, 1 (2012).

\bibitem{data}
P.~A.~R.~Ade, et al. (Planck Collaboration), Astron.\ Astrophys.\ {\bf 571}, A16 (2014); P.~A.~R.~Ade, et al. (Planck Collaboration), Astron.\ Astrophys.\ {\bf 594}, A13 (2016).

\bibitem{des}
M.~A.~Troxel, et~al. (DES Collaboration), arXiv:1708.01538[astro-ph.CO].

\bibitem{riess}
A.~G.~Riess, et~al. (Supernova Search Team), Astron.\ J. {\bf 116}, 1009 (1998). 

\bibitem{perlmutter}
S.~Perlmutter, et~al. (Supernova Cosmology Project), Astrophys.\ J. {\bf 517}, 565 (1999).

\bibitem{general}
R.~R.~Caldwell, Phys.\ Lett.\ B {\bf 545}, 23 (2002); B.~Feng, X.~Wang and X.~Zhang, Phys.\ Lett.\ B {\bf 607}, 35 (2005).

\bibitem{libro}
S.~Tsujikawa, ``Dark Energy: Investigation and Modeling''. Dark Matter and Dark Energy. Astrophysics and Space Science Library, {\bf 370}, 331-402, Springer (2011).

\bibitem{pavon1}
D.~Pavon, J.\ Phys.\ A {\bf 40}, 6865 (2007).

\bibitem{od1}
S.~Nojiri and S.~D.~Odintsov, Gen.\ Rel.\ Grav. {\bf 38}, 1285 (2006).

\bibitem{od2}
S.~Nojiri and S.~D.~Odintsov, Eur.\ Phys.\ J.\ C {\bf 77}, 528 (2017).

\bibitem{int}
S.~del~Campo, J.~C.~Fabris, R.~Herrera and W.~Zimdahl, Phys.\ Rev.\ D {\bf 83}, 123006 (2011); J.~B.~Jim\'enez, D.~Rubiera-Garcia, D.~S\'aez-G\'omez and V.~Salzano, Phys.\ Rev.\ D {\bf 94}, 123520 (2016); F.~Ar\'evalo, P.~Cifuentes, S.~Lepe and F.~Pe\~na, Astrophys.\ Space\ Sci. {\bf 352}, 899 (2014); A.~A.~Costa, R.~C.~G.~Landim, B.~Wang and E.~Abdalla, arXiv:1803.06944[astro-ph.CO]; C.~Li and Yi-Fu~Cai, arXiv:1804.04816[astro-ph.CO]; E.~Aydiner, Scientific Reports {\bf 8}, 721 (2018); M.~Cruz, S.~Lepe and F.~Pe\~na, Phys.\ Rev.\ D {\bf 92}, 123511 (2015).

\bibitem{chemical}
M.~Cruz, S.~Lepe and S.~D.~Odintsov, Phys.\ Rev.\ D {\bf 98}, 083515 (2018).

\bibitem{li}
M.~Li, Phys.\ Lett.\ B {\bf 603}, 1 (2004).

\bibitem{joel}
N.~Cruz, S.~Lepe, F.~Pe\~na and J.~Saavedra, Phys.\ Lett.\ B {\bf 669}, 271 (2008).

\bibitem{miguel}
M.~Cruz and S.~Lepe, Class.\ Quantum\ Grav. {\bf 35}, 155013 (2018).

\bibitem{curv1}
O.~Farooq, D.~Mania and B.~Ratra, Astrophys.\ Space\ Sci. {\bf 357}, 11 (2015).

\bibitem{curv2}
O.~Farooq, F.~Madiyar, S.~Crandall and B.~Ratra, Astrophys.\ J. {\bf 835}, 26 (2017). 

\bibitem{bak}
D.~Bak and S.~J.~Rey, Class.\ Quantum\ Grav. {\bf 17}, L83 (2000).

\bibitem{pavon}
N.~Radicella and D.~Pav\'on, J.\ Cosmol.\ Astropart.\ Phys. {\bf 10}, 005 (2010).

\bibitem{velten}
H.~E.~S.~Velten, R.~F.~vom~Marttens and W.~Zimdahl, Eur.\ Phys.\ J.\ C {\bf 74}, 3160 (2014).

\bibitem{datadecel}
A.~A.~Mamon, Mod.\ Phys.\ Lett.\ A {\bf 33}, 1850056 (2018). 
\end{thebibliography}
\end{document}